\documentclass[epsfig,useAMS,usenatbib]{mn2e}
\usepackage{epsfig,graphics}

\def\grtsim{\mathrel{\hbox{\rlap{\hbox{\lower2pt\hbox{$\sim$}}}\raise2pt\hbox{$>$}}}}
\def\lesssim{\mathrel{\hbox{\rlap{\hbox{\lower2pt\hbox{$\sim$}}}\raise2pt\hbox{$<$}}}}
\newcommand{\be}{\begin{equation}}
\newcommand{\ee}{\end{equation}}
\title[Galaxy clusters at $0.6 < z < 1.4$ in the UKIDSS-UDS]{Galaxy
  clusters at {$\bf 0.6 < z < 1.4$} in the UKIDSS Ultra Deep Survey Early Data Release}
\author[van Breukelen et al.]{C. van Breukelen,$^{1}$
  L. Clewley,$^{1}$ D. G. Bonfield,$^{1}$ S. Rawlings,$^{1}$ M. J. Jarvis,$^{1}$ 
\newauthor J. M. Barr,$^{1}$ S. Foucaud,$^{2}$ O. Almaini,$^{2}$ M. Cirasuolo,$^{3}$ G. Dalton,$^{1}$ 
\newauthor J. S. Dunlop,$^{3}$ A. C. Edge,$^{4}$ P. Hirst,$^{5}$ R. J. McLure,$^{3}$
M. J. Page,$^{6}$ 
\newauthor K. Sekiguchi,$^{7}$ C. Simpson,$^{8}$ I. Smail,$^{4}$ M. G. Watson$^{9}$\\
$^{1}$Astrophysics, Department of Physics, Keble Road, Oxford, OX1 3RH, UK (E-mail: cvb@astro.ox.ac.uk)\\
$^{2}$School of Physics and Astronomy, University of Nottingham, University Park, Nottingham, NG7 2RD, UK\\
$^{3}$Scottish Universities Physics Alliance, Institute for Astronomy, University of Edinburgh, Royal
Observatory, Edinburgh EH9 3HJ, UK\\
$^{4}$Institute for Computational Cosmology, Department of Physics,
Durham University, Durham DH1 3LE, UK\\
$^{5}$Joint Astronomy Centre, 660 N. A'ohoku Place, University Park,
Hilo, Hawaii 96720, U.S.A.\\
$^{6}$Mullard Space Science Laboratory, University College London, Holmbury St. Mary, Dorking, Surrey RH5 6NT, UK\\
$^{7}$Subaru Telescope, National Astronomical Observatory of Japan, 650 N. A'ohoku Place, Hilo, Hawaii 96720, USA\\
$^{8}$Astrophysics Research Institute, Liverpool John Moores
University, Twelve Quays House, Egerton Wharf, Birkenhead CH41 1LD, UK\\
$^{9}$X-ray Astronomy Group, Department of Physics and Astronomy,
University of Leicester, Leicester LE1 7RH, UK}
\date{Released 2006 Xxxxx XX}

\pagerange{\pageref{firstpage}--\pageref{lastpage}} \pubyear{2002}

\def\LaTeX{L\kern-.36em\raise.3ex\hbox{a}\kern-.15em
    T\kern-.1667em\lower.7ex\hbox{E}\kern-.125emX}

\begin{document}
\label{firstpage}
\maketitle
\begin{abstract}
We present the first cluster catalogue extracted from the UKIRT
Infrared Deep Sky Survey Early Data Release. The catalogue is created
using UKIDSS Ultra Deep Survey infrared $J$ and $K$ data combined with
3.6 $\mu$m and 4.5 $\mu$m Spitzer bands and optical $BVRi'z'$ imaging from the
Subaru Telescope over 0.5 square degrees in the Subaru XMM-Newton Deep
Field. We have created a new cluster-detection algorithm, based
on the Friends-Of-Friends and Voronoi Tessellation methods, which
utilises probability distribution functions derived from a photometric
redshift analysis. We employ mock catalogues to understand the
selection effects and contamination associated with the algorithm. The
cluster catalogue contains 13 clusters at redshifts 0.61 $\leq z \leq$
1.39 with luminosities $10 {\rm L^*} \lesssim L_{\rm tot} \lesssim 50
\rm L^*$, corresponding to masses $5 \times 10^{13} ~ {\rm M_{\odot}}
\lesssim M_{cluster} \lesssim 3 \times 10^{14} ~ \rm M_{\odot}$ for
$\frac{M/\rm M_{\odot}}{L/\rm L_{\odot}} = 75h$. The measured sky
surface density of $\sim 10 ~ \rm deg^{-2}$ for high-redshift
($z=0.5-1.5$), massive ($>10^{14} ~ \rm M_{\odot}$) clusters is
precisely in line with theoretical predictions presented by Kneissl et al.\
(2001).
 
\end{abstract}


\section{Introduction}
\vspace{-0.2cm} Distant galaxy clusters are used in a wide range of
cosmological and astrophysical contexts. For instance, cluster studies
illuminate how dark matter haloes collapse and large-scale structure
forms and evolves. Comparisons of the evolution of the cluster
mass-function to that predicted by $N$-body simulations (Evrard et
al. 2002) or the Press-Schechter formalism (Press \& Schechter 1974)
and its variants place strong constraints on cosmological parameters
such as the mass density ($\Omega_{\rm M}$) and the amplitude of the
mass fluctuations in the early Universe ($\sigma_8$) (e.g. Eke et
al. 1998). Further, high-redshift clusters can help constrain
feedback processes caused by star-formation and Active Galactic Nuclei
(e.g. Silk \& Rees 1998). Unfortunately few $z>1$ clusters have been
identified, and the vast majority of these are from X-ray surveys
(e.g. using XMM-Newton; Stanford et al. 2006).

\vspace{-0.05cm}
X-ray selected cluster surveys (e.g. Mullis et al. 2005) have
succeeded in finding clusters with luminous intra-cluster media. Some
follow-up observations exploiting the Sunyaev-Zel'dovich (SZ) effect
have been successful (e.g. Jones et al. 2005) and comprehensive SZ
surveys are underway (Kneissl et al. 2001). Optical selection of
clusters is complementary to these studies (Gilbank et al. 2004) as it
is based on other cluster properties, e.g. optical vs. X-ray luminosity or
the SZ decrement of the cluster gas. An added advantage of using
optical data is the possibility of employing a photometric redshift
analysis to derive the distance to the clusters and gain a
three-dimensional perspective. Optical cluster searches however have
been stymied at high redshifts by the fact that the 4000\AA\ break
(ubiquitous for the early-type, red galaxies dominant in clusters)
shifts out of the $I$-band into the near-infrared. Recent developments
have seen the advent of large near-infrared cameras like the Wide
Field Infrared Camera (WFCAM) on the United Kingdom Infrared Telescope
(UKIRT). WFCAM is now undertaking the UKIRT Infrared Deep Sky Survey
(UKIDSS, Lawrence et al. 2006).

In this letter we use the UKIDSS Ultra Deep Survey (UDS) Early Data
Release (EDR, Dye et al. 2006) to find clusters in the redshift range
$0.5 \leq z \leq 1.5$. The clusters are found by applying adaptations
of two cluster selection methods: the Voronoi Tessellation technique
(Ebeling \& Wiedenmann 1993) and the Friends-Of-Friends method (Huchra
\& Geller 1982). We describe the data and the determination of
photometric redshifts in \S2, and the cluster-detection algorithm and
simulations using mock catalogues in \S3. The cluster catalogue is
presented in \S4 and in \S5 we summarize our conclusions. Throughout
this letter we assume $H_{\circ}=71~ {\rm km~s^{-1}Mpc^{-1}}$,
$\Omega_ {\rm M}=0.27$, and $\Omega_ {\Lambda}=0.73$. All magnitudes
quoted are in the AB-system.

\section{Data and Photometric Redshifts}
We use three sources of data: near-infrared $J$ and $K$ data from the
UDS EDR (Foucaud et al. 2006); 3.6$\mu$m and 4.5$\mu$m bands from the
Spitzer Wide-area InfraRed Extragalactic survey (SWIRE, Lonsdale et
al. 2005); and optical $BVRi'z'$ Subaru data over the Subaru
XMM-Newton Deep Field (SXDF, Furusawa et al. in prep.). The area of
our cluster survey is 0.5 sq. deg. with 34.2$^{\circ} <$ RA $<
34.8^{\circ}$ and $-$5.4$^{\circ} <$ Dec. $< -4.6.^{\circ}$ (02:16:48.00
$<$ RA $<$ 02:19:13.20, $-$05:25:12.0 $<$ Dec $<$ $-$04:38:17.1).  The
galaxy catalogue includes the objects with a detection in $i'$, $J$
and $K$ and to exclude stars we impose a criterion of SExtractor
stellarity index $<$ 0.8 in $i'$ and $K$ (e.g. Bertin \& Arnouts
1996). The depth of the catalogue is limited by the UDS EDR 5$\sigma$
magnitude limits of $K_{\rm AB,lim} = J_{\rm AB,lim} = 22.5$. 
We derive photometric redshifts by fitting a spectral energy
distribution (SED) template to each object's photometry using the
\emph{Hyperz} code (Bolzonella et al 2000). The galaxy templates are
generated with the stellar population synthesis code GALAXEV (Bruzual
\& Charlot 2003) and cover a range of different star formation rates
with timescales $\tau$ from 0.1 to 30 Gyr. We adopt a flat prior for
galaxy luminosity up to a maximum of $L = 10 \rm L^*$ (assuming a
passively evolving elliptical galaxy) in the observed $K$-band and calculate
the marginalised posterior redshift probability functions. Our
photometric redshift catalogue comprises 19300 objects in the range
$0.1 \leq z \leq 2.0$.

\subsection{Reliability of the photometric redshifts}
To test the reliability of the photometric redshifts we run
\emph{Hyperz} with the same templates on spectroscopically identified
objects in the SXDF (Yamada et al. 2005; Simpson et al. 2006).
The spectroscopic and photometric redshifts are consistent, with a
photometric redshift error of $\sigma_z = 0.08$ over $0.5 \leq z \leq
1.5$ and no obvious systematic offset. However, an examination of the
number of galaxies in the photometric-redshift catalogue as a function
of redshift, $N(z)$, reveals spikes in the redshift distribution. The
shape of the distribution is constant across the field which indicates
the spikes are not caused by large-scale structure but by aliasing
effects inherent to photometric redshift methods employing many more
templates than data points (e.g. Rowan-Robinson 2003). A serious
concern is the detection of spurious clusters due to this focussing in
redshift. We therefore create a second redshift catalogue using four
empirically derived SEDs from Coleman, Wu \& Weedman (1980), supplied
with the \emph{Hyperz} code. The new galaxy redshift probability
functions are broader; the spikes disappear yielding a smoother $N(z)$
distribution. However, a comparison of the spectroscopic redshifts
reveal that the redshift is systematically underestimated at $z
\grtsim 0.5$, with an offset of $\Delta z = 0.15$ at $z \sim 1$.

In summary, an examination of the accuracy of the photometric
redshifts reveals a sensitivity to the templates used in the
photometric redshift estimation. Redshifts quoted in this letter use
the Bruzual \& Charlot templates as the spectroscopic sample show
these to be most accurate. However to avoid false cluster detections
due to the spikiness in the $N(z)$ distribution we limit ourselves in
this letter to the clusters that are isolated in both photometric
redshift catalogues and are therefore considered to be robust
detections.

\section{The Algorithm}
The cluster-detection algorithm is described in detail in van
Breukelen et al. (in prep.). Here we present only a brief outline of
the procedures we employ.

Two common problems of optically selected cluster samples are the
detection of spurious clusters, and the chance projection of fore- and
background galaxies into the real clusters. This is caused by
selection biases inherent to any detection algorithm and the fact that
photometric redshift probability functions (z-PDFs) can take any form,
exhibiting errors that may be many times larger than the redshift
range of the cluster. To reduce these difficulties we (i) use two
substantially different cluster-detection methods to minimize the
number of false detections, and (ii) utilize the full z-PDFs rather
than the best redshift estimate with an associated error. For each
method we sample the z-PDFs to create 500 Monte-Carlo (MC)
realisations of the 3-D galaxy distribution. These are divided in
slices of $\Delta z = 0.05$\footnote{When the analysis was re-run with
the alternative photometric redshift catalogue (Sec 2.1) this was
changed to $\Delta z = 0.1$ to account for the broader z-PDFs.}  in
which the cluster candidates are identified. Since the error on the
photometric redshift is larger than the width of the redshift slices,
each cluster-candidate is typically found in several adjoining
slices. We determine the final cluster-redshift by taking the average
of the redshifts at which the cluster occurs, weighted by the
corresponding number of MC-realisations. We assign a `reliability
factor' $F$ to each of the clusters which is the total fraction of
MC-realisations in which the cluster is detected.
The algorithm uses the Voronoi Tessellation technique (VT), and the
Friends-Of-Friends method (FOF) to detect cluster candidates in the
redshift slices. One of the principal advantages of the VT method is
that it is relatively unbiased as it does not look for a particular
source geometry (e.g.  Ramella et al. 2001).  The field of galaxies is
divided into Voronoi Cells, each containing one object: the
nucleus. The reciprocal of the area of the VT cells translates to a
local density. Overdense regions in the plane are found by fitting a
function (see Kiang 1966) to the density distribution of all VT cells
in the field; cluster candidates are the groups of cells of a
significantly higher density than the mean background density. The
Friends-Of-Friends (FOF) algorithm groups galaxies with a smaller
separation than a projected linking distance $D_{\rm link}$
(`friends'). When spectroscopic data is available, the `friends' are
also subject to a linking velocity $\Delta V \leq V_{\rm link}$
(e.g. Tucker et al., 2002). With photometric redshifts only galaxies
within the same redshift slice are linked (see Botzler et
al. 2004). If a group comprises a number of galaxies larger than
$n_{\rm min}$ it is considered a cluster candidate.

We create mock catalogues that mimic the real data to quantify the
various selection effects and expected contaminants. These simulations
also lead to estimates of the total stellar luminosity of the clusters
found in the real data.

\subsection{Simulated Cluster Catalogues}
We create catalogues with a galaxy distribution randomly
placed\footnote{We neglect clustering of both the background and the
clusters.} in the field with $0.1 \leq z \leq 2.0$. The galaxy
luminosities and number densities are determined by the $K$-band
luminosity function of Cole et al. (2001), with the simplifying
assumption of passive evolution with formation redshift
$z_{form}=10$. A detection limit of $K < 22.5$ is imposed to match the
5-$\sigma$ limit of the UDS EDR data. The number of galaxies as a
function of magnitude in each mock catalogue is entirely consistent
with the number counts in the UDS catalogue to this limit. We
superimpose simulated clusters of total mass $M = 0.5, 1.0, 2.0, 10,
20 \times 10^{14}~\rm M_{\odot}$, assuming $\frac{M/\rm
M_{\odot}}{L/\rm L_{\odot}} = 75h$ (Rines et al. 2001) and applying a
cluster $K-$band luminosity function (Lin et al. 2004). The galaxies
are spatially distributed within the cluster according to an NFW
profile (Navarro, Frenk \& White, 1997) with a cut-off radius of 1
Mpc. For the simulated clusters of mass $> 1.0 \times 10^{15}~\rm
M_{\odot}$ this is smaller than the virial radius; however we do not
find such high mass clusters in our data (see Section 4) and the
effect of the cut-off is therefore negligible.  The clusters are
placed at redshifts $z = 0.1, 0.2, ..., 2.0$. Each combination of mass
and redshift is represented in ten randomly created catalogues. We
offset the redshift of all galaxies randomly with a factor equal to
the errors in the real data. Each galaxy receives a z-PDF taken from
the data by selecting an object with corresponding redshift.

The VT and FOF methods each have two free parameters. For FOF these
are the linking distance in proper coordinates, $D_{\rm link}$, and
the minimum number of galaxies in a cluster, $n_{\rm min}$. Guided by
Botzler et al. (2004) we experiment with values between 0.125 Mpc
$\leq D_{\rm link} \leq$ 0.175 Mpc, and $3 \leq n_{\rm min} \leq
5$. For VT the parameters are the maximum probability of an
overdensity being a background fluctuation, $p_{\rm bg}$, and the
lower limit on the cell density, $f_{\rm min}$. We follow the method
of Ebeling \& Wiedenmann (1993) and set $p_{\rm bg}$ to 10\%. For
$f_{\rm min}$ we try values of $1.2-2.2$, where $f = 1.0$ is the mean
cell density of the field. We use the parameters that optimize the
algorithm's performance: $D_{\rm link} = 0.175$ Mpc, $n_{\rm min} =
5$, and $f_{\rm min} = 1.74$. Since both methods use different
measures to isolate clusters (galaxy density in VT versus separation
in FOF) the false detections in both do not coincide. Therefore by
cross-correlating the output of the two methods the spurious sources
due to biases in the algorithms disappear, which reduces the
contamination to chance galaxy groupings. There is no obvious bias due
to cluster morphology (for a full treatment of biases and contanimation
see van Breukelen et al in prep.).

\begin{figure*}
\epsfig{figure=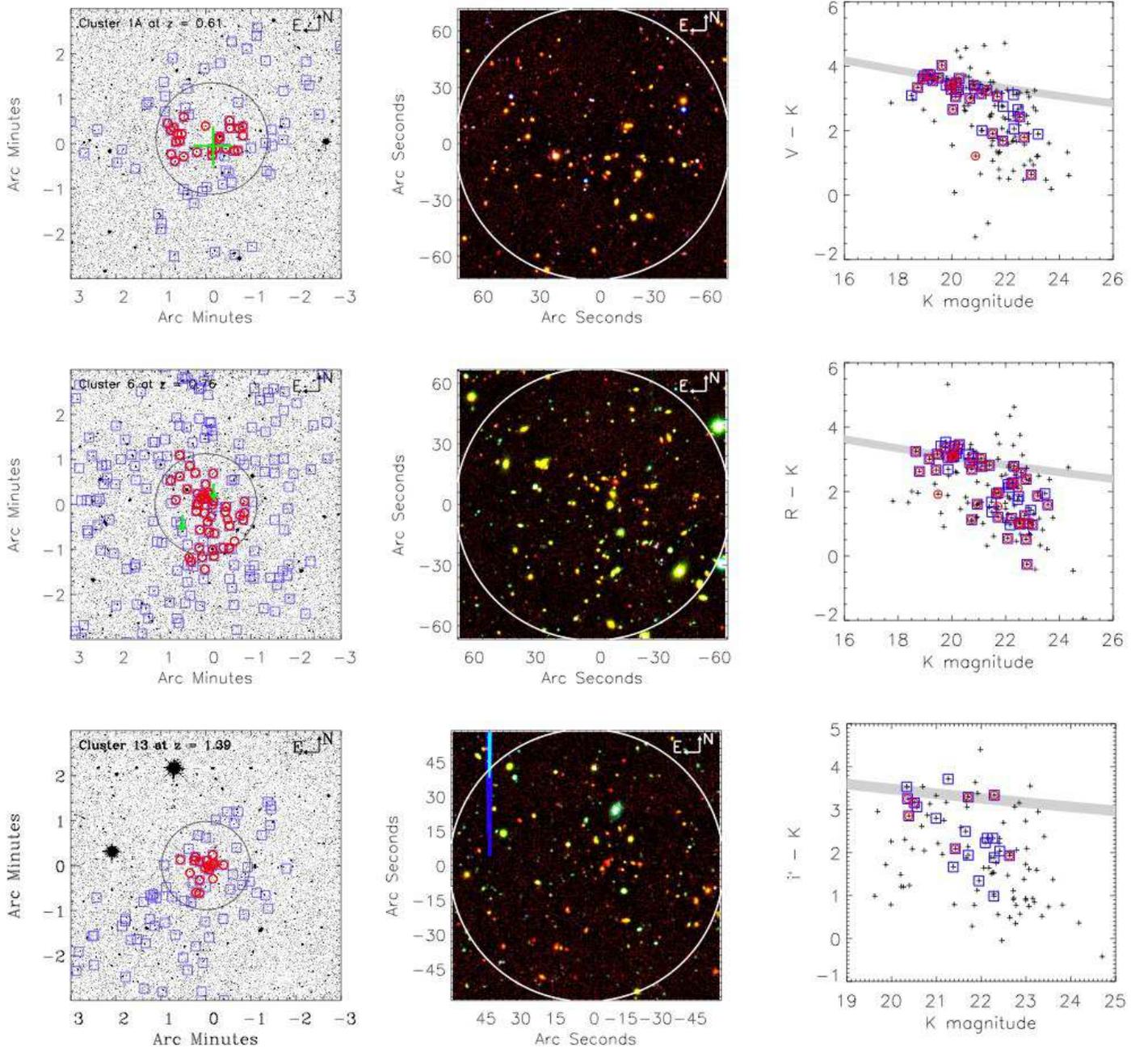, height=175mm}
\caption{\small Clusters 1, 6, and 13 spanning the redshift range of
  our detections. {\em Left:} $K$-band images; the large circle shows
  a 1 Mpc region around the cluster; the blue squares and red circles
  are cluster members as given by VT and FOF respectively. ID 1: the
  green cross denotes an extended X-ray source. ID 6: the green arrows
  point out galaxies with $z_{\rm spec} = 0.87$. {\em Middle:} $Bz'K$
  images of the central 1 Mpc region. {\em Right:} colour-magnitude
  plots of the clusters: the colour is the reddest filter shortward of
  4000\AA\ (restframe) minus $K$. The crosses are all galaxies within
  the central 1 Mpc region, otherwise the symbols are the same as in
  the left-hand panel. The grey band is the modelled red sequence.}
\label{colour_plots}
\end{figure*}

We compare the recovered cluster galaxies to the input galaxies of the
mock clusters. VT tends to include all galaxies in a large area around
the cluster core (see for example the left-hand panel of
Fig.~\ref{colour_plots}). The number of recovered cluster members,
$N_{\rm gal}$, in any cluster is sensitive to the local field
density. By contrast, the galaxy members recovered by FOF are more
centrally concentrated; the total number per cluster is consistent
throughout the random realisations of the catalogues. Thus we use both
methods to detect the clusters and only FOF to calculate $N_{\rm
gal}$. This is done by taking all galaxies that occur in the cluster
at $>15\%$ of the MC-realisations in which the cluster itself is
detected. The galaxies that appear in a smaller fraction of
MC-realisations are very likely to be interlopers from different
redshifts.
Calculating $N_{\rm gal}$ for all cluster-masses at all redshifts
yields functions of $N_{\rm gal}$ vs. $z$ for constant mass or total
luminosity. At $z > 1.5$ only clusters of masses $M \geq 1.0 \times
10^{15}~\rm M_{\odot}$ or $L > 170\rm L^*$ can be detected with $\sim$
50\% completeness. At redshifts of $z < 0.5$ our field of view is
$\lesssim$ 10 times the typical cluster size of 1 Mpc. Hence we limit
our redshift range to $0.5 \leq z \leq 1.5$, allowing the detection of
clusters of luminosity $L \grtsim 10 \rm L^*$ at $z \sim 0.5$ to $L
\grtsim 170 \rm L^*$ at $z \sim 1.5$.
\vspace{-0.5cm}

\section{Results: The Cluster Catalogue}

\begin{table*}
\vspace{-0.4cm}
{\small
\begin{tabular}{lccccccccl}
\hline
\hline
\noalign{\smallskip} 
\multicolumn{1}{c}{ID} & 
\multicolumn{1}{c}{RA} &  
\multicolumn{1}{c}{Dec.} &  
\multicolumn{1}{c}{$z_{\rm phot}$} & 
\multicolumn{1}{c}{$N_{\rm gal}$}  &
\multicolumn{1}{c}{$L_{\rm tot}$} &  
\multicolumn{1}{c}{Mass} &  
\multicolumn{1}{c}{$F_{\rm fof}$} &  
\multicolumn{1}{c}{$F_{\rm vt}$} &
\multicolumn{1}{c}{$z_{\rm spec}$} \\ 
\multicolumn{1}{c}{} &
\multicolumn{1}{c}{[h m   s]} &   
\multicolumn{1}{c}{[$^{\circ}$  $^{\prime}$  $^{\prime\prime}$]} &   
\multicolumn{1}{c}{} &
\multicolumn{1}{c}{} &
\multicolumn{1}{c}{[$\rm L^*$]} &
\multicolumn{1}{c}{[10$^{14}~\rm M_{\odot}$]} &
\multicolumn{1}{c}{} &
\multicolumn{1}{c}{} &
\multicolumn{1}{c}{} \\
\noalign{\smallskip} 
\hline 
\noalign{\smallskip} 
1A*               &  02 17 35.3  &  -05 13 16  & 0.61  $\pm$ 0.05 & 24  & 19   & 1.1  & 0.8  & 0.5  & 0.65$^\dag$\\
1B                &  02 17 31.5  &  -05 10 59  & 0.65  $\pm$ 0.07 & 10  & 10   & 0.5  & 0.4  & 0.4  & -\\
2A                &  02 16 59.9  &  -05 10 39  & 0.64  $\pm$ 0.02 & 11  & 10   & 0.6  & 0.7  & 0.2  & -\\
2B                &  02 16 51.7  &  -05 12 15  & 0.71  $\pm$ 0.05 & 17  & 17   & 1.0  & 0.7  & 0.5  & -\\
3                 &  02 18 34.9  &  -04 58 05  & 0.66  $\pm$ 0.07 & 23  & 19   & 1.1  & 1.0  & 0.7  & -\\
4                 &  02 18 09.0  &  -05 21 54  & 0.67  $\pm$ 0.03 & 19  & 17   & 1.0  & 0.5  & 0.3  & -\\
5                 &  02 18 00.4  &  -04 42 58  & 0.71  $\pm$ 0.03 &  8  & 10   & 0.5  & 0.3  & 0.2  & -\\
6*                &  02 18 32.7  &  -05 01 04  & 0.76  $\pm$ 0.12 & 36  & 46   & 2.7  & 1.0  & 0.8  & 0.87$^{\ddag}$\\
7                 &  02 19 03.5  &  -04 42 33  & 0.78  $\pm$ 0.06 & 17  & 18   & 1.0  & 0.7  & 0.3  & -\\
8                 &  02 17 56.4  &  -05 02 46  & 0.79  $\pm$ 0.07 & 11  & 11   & 0.6  & 0.2  & 0.2  & -\\
9                 &  02 17 21.4  &  -05 11 30  & 0.80  $\pm$ 0.06 & 15  & 17   & 1.0  & 0.2  & 0.6  & 0.93$^{\dag\dag}$\\
10                &  02 18 03.2  &  -05 00 01  & 0.94  $\pm$ 0.14 & 17  & 36   & 2.1  & 0.4  & 0.8  & -$^{\ddag\ddag}$\\
11A               &  02 18 06.4  &  -05 03 25  & 0.95  $\pm$ 0.11 & 24  & 48   & 2.8  & 0.6  & 0.8  & 1.06$^{\ddag}$\\
11B               &  02 18 15.5  &  -05 02 50  & 0.98  $\pm$ 0.09 & 12  & 18   & 1.0  & 0.3  & 0.4  & 0.92$^{\ddag}$\\
12                &  02 17 07.4  &  -04 46 44  & 1.01  $\pm$ 0.06 & 16  & 39   & 2.3  & 0.3  & 0.6  & 1.102$^{\dag\dag\dag}$\\
13*               &  02 18 10.5  &  -05 01 05  & 1.39  $\pm$ 0.07 &  7  & 41   & 2.4  & 0.2  & 0.6  & -\\
\hline
14                &  02 18 37.3  &  -04 48 50  & 0.77  $\pm$ 0.07 & 11  & 11   & 0.6  & 0.4  & 0.3  & -\\
\noalign{\smallskip} 
\hline
\hline    
\end{tabular}

}
\vspace{-0.3cm}
\caption{\small The cluster catalogue ordered by redshift. The
  position (J2000) is the centroid of the cluster galaxies. $N_{\rm
  gal}$ is the number of cluster galaxies detected by FOF with $K <
  22.5;$ $L_{\rm tot}$ is the estimated total luminosity of the
  cluster and mass the inferred total cluster mass; $F_{\rm fof}$ and
  $F_{\rm vt}$ are the reliability factors given respectively by FOF
  and VT. The possible spectroscopic redshift is given as $z_{\rm
  spec}$; clusters marked with a star [*] are shown in
  Fig.~\ref{colour_plots}. All clusters above the line are robust; ID
  14 was not recovered with the alternative photometric redshift
  catalogue. \dag~ Extended X-ray-detected cluster (Kolokotronis et
  al. 2006); $z _{\rm spec}$ from Geach et al. (in prep). \ddag~ Two
  galaxies in each of the clusters are found at this redshift (Yamada
  et al. 2005). \dag\dag~ Four galaxies in the cluster are found at
  this redshift (Yamada et al. 2005). \ddag\ddag~ Potential cluster
  galaxies have $z_{\rm spec} = $0.874, 0.96, and 1.095;
  super-position effects may be exaggerating the richness of
  this cluster. \dag\dag\dag~ Redshift of a QSO possibly associated
  with the cluster (Sharp et al. 2002).}
\label{table_clusters}
\end{table*}

Our final cluster catalogue comprises all clusters found by both VT
and FOF with a reliability factor of $F > 0.2$ to exclude false
detections. We calculate the redshift by averaging the cluster
redshift given by the VT and FOF methods. This results in 14 clusters
at $0.61 \leq z \leq 1.39$; Table~\ref{table_clusters} lists all the
clusters with their positions and properties (full details of the
cluster galaxies are available on request). The redshift error given
in Table~\ref{table_clusters} combines the errors given by VT and FOF
and reflects the range of redshift slices in which the cluster is
detected. Although the cluster-finding method is sensitive to clusters
down to a luminosity of $\sim 10 \rm L^*$ we are limited by the
relatively large systematic errors from the photometric redshifts. The
13 clusters `above the line' in Table~\ref{table_clusters} were also
detected in our alternative photometric redshift catalogue (see
Section 2.1), although sometimes only at the $F_{\rm fof}$ and $F_{\rm
vt} > 0.1$ level; thus these are judged robust detections. Only one
cluster (`below the line' in Table~\ref{table_clusters}) was not
recovered with the alternative photometric redshift catalogue.
Fig.~\ref{colour_plots} shows three cluster examples. On the left a
$K$-band image, a combined $Bz'K$ image in the middle, and on the
right a colour-magnitude (CM) diagram. For illustration a modelled red
sequence is overplotted on the CM-diagram (using GALAXEV elliptical
templates of $z_{\rm form}$ = 10, $\tau = 0.5$ Gyr and a slope derived
from Kodama \& Arimoto 1997). A clear red sequence can be seen for the
top two clusters; the third is unclear because of the small number of
cluster galaxies, although four of the seven galaxies detected in both
VT and FOF lie near the predicted red sequence. 

Three clusters (no. 1, 2, 11) consist of two concentrations of
galaxies separated by $\lesssim$ 1 Mpc in the FOF method, but as one
cluster in VT: these could be merging clusters.
We determine $N_{\rm gal}$ with $K < 22.5$ (corresponding to the
completeness limit) in the same way as for our simulated clusters (see
Section 3.1); this allows us to derive an approximate total luminosity
to the cluster by interpolating between the lines of constant total
luminosity in the $N_{\rm gal}-z$ plane found in our simulations. We
find our clusters span the range of $10 {\rm L^*} \lesssim L_{\rm tot}
\lesssim 50 \rm L^*$; assuming $\frac{M/\rm M_{\odot}}{L/\rm
L_{\odot}} = 75h$ (Rines et al. 2001) this yields $0.5 \times
10^{14}~{\rm M_{\odot}} \lesssim M_{\rm cluster} \lesssim 3 \times
10^{14}~\rm M_{\odot}$.

\section{Concluding Remarks}
Our cluster catalogue comprises 13 clusters at redshifts 0.61 $\leq z
\leq$ 1.39 with estimated luminosities $10 {\rm L^*} \lesssim L_{\rm
tot} \lesssim 50 \rm L^*$, corresponding to masses $5 \times
10^{13}~{\rm M_{\odot}} \lesssim M_{\rm cluster} \lesssim 3 \times
10^{14}~\rm M_{\odot}$. Considering just the clusters with total mass
estimates significantly $>10^{14} ~ \rm M_{\odot}$ (6, 10, 11, 12, 13),
this represents a sky surface density of $\sim 10 ~\rm deg^{-2}$ for
high-redshift ($z=0.5-1.5$) clusters. This is in quantitative
agreement with the predictions of low-density ($\Omega_{\rm M} \approx
0.3$) cosmological models, e.g.\ those presented by Kneissl et al.\
(2001). Comparing with their Figs.\ 5 \& 8, we see that, within the
obvious limitations of small-number statistics, the real clusters in
the SXDF have the same abundance, redshift and estimated mass
distributions as are predicted by theory.

Clearly spectroscopic observations are essential to
confirm the reality of the clusters. From spectroscopic redshifts in the
literature (see Table~\ref{table_clusters}), we have tentative
confirmation of the photometric redshifts for 6 out of 9 clusters
at $z > 0.75$. In the near future near-infrared multi-object
spectrometers on 8-metre class telescopes will provide opportunities
for spectroscopic follow-up of high-redshift clusters.

As we note only one of our cluster candidates is associated with a
published extended X-ray-detected cluster (Table~\ref{table_clusters}),
we plan to exploit existing deep XMM-Newton data (Ueda et al. in
prep.) and upcoming SZ surveys to determine the gas contents of these
clusters.

\vspace{-0.5cm}
\section{Acknowledgements}
This work is based on data obtained from UKIDSS and the Spitzer Space Telescope.
UKIRT is operated by the Joint Astronomy Centre on
behalf of PPARC. We are grateful to the staff at UKIRT, Subaru and
Spitzer for making these observations possible. We acknowledge
CASU in Cambridge and WFAU in Edinburgh for processing the UKIDSS
data. CvB, LC, DGB, SR, MJJ, SF, CS, and MC acknowledge funding from
PPARC. OA, RJM, and IS acknowledge the support of the Royal Society.

\end{document}